# Observation of electromagnetically induced Talbot effect in an atomic system

Zhaoyang Zhang[1,2], Xing Liu[2], Dan Zhang[1,2], Jiteng Sheng[3], Yiqi Zhang[2], Yanpeng Zhang[2*] and Min Xiao[1,4*]

[1]*Department of Physics, University of Arkansas, Fayetteville, Arkansas 72701, USA*

[2]*Key Laboratory for Physical Electronics and Devices of the Ministry of Education & Shaanxi Key Lab of Information Photonic Technique, Xi'an Jiaotong University, Xi'an 710049, China*

[3]*State Key Laboratory of Precision Spectroscopy, East China Normal University, Shanghai 200062, P. R. China*

[4]*National Laboratory of Solid State Microstructures and School of Physics, Nanjing University, Nanjing 210093, China*

[*]*Corresponding author: ypzhang@mail.xjtu.edu.cn, mxiao@uark.edu*

## Abstract

We experimentally demonstrate the Talbot effect resulting from the repeated self-reconstruction of a spatially intensity-modulated probe field under the Fresnel near-field regime. By launching the probe beam into an optically induced atomic lattice (established by interfering two coupling fields) inside a rubidium vapor cell, we can obtain an diffracted probe beam pattern from an formed electromagnetically induced grating (EIG) in a three-level Λ-type Doppler-free atomic configuration with the assistance of electromagnetically induced transparency (EIT). The EIG-based diffraction pattern repeats itself at the planes of integer multiple Talbot lengths, which agrees well with the theoretical prediction [*Appl. Phys. Lett.*, **98**, 081108 (2011)]. In addition, fractional EIT-induced Talbot effect was also investigated. Such experimentally demonstrated EIT Talbot effect in a coherently-prepared atomic system may pave a way for lensless and nondestructive imaging of ultracold atoms and molecules, and further demonstrating nonlinear/quantum beam dynamical features predicted for established periodic optical systems.

The conventional Talbot effect characterized as a self-imaging or lensless imaging phenomenon was first implemented by launching a very weak white light into a Fraunhofer diffraction grating and observing the images of the same periodic structure at certain periodical distances with integer multiples of Talbot length [1, 2]. The wide practicability and simplicity of the self-imaging processes have inspired continuous and in-depth researches on the Talbot effect in various systems.[3] Since the invention of coherent light sources, studies of this near-field diffraction phenomenon have made great progresses not only in optics, such as optical measurements [4] and optical computing [5], but also in a variety of new areas including waveguide arrays [6], parity-time symmetric optics [7], X-ray diffraction [8], Bose–Einstein condensates [9], second-harmonic generation [10], quantum optics [11, 12] and the recently proposed Airy-Talbot effect [13-15]. In 2011, the Talbot effect based on an electromagnetically induced grating (EIG) [16] in an atomic medium was theoretically proposed to produce the self-imaging of ultracold atoms or molecules without using any sophisticated optical components [17]. Compared with the commonly used on/off-resonant absorption imaging methods [18, 19], such predicted electromagnetically induced Talbot effect (EITE) can partially maintain the advantages such as nondestructive detection and overcome the shortcoming of using complicated optical setup for in-situ imaging, which may make the new imaging method be an alternative approach for directly observing ultracold atoms or molecules in a cloud. Furthermore, based on such proposed EITE, nonlocal atomic imaging schemes, relied on the second-order two-photon EITE [20] and the second-order self-imaging with parametrically amplified four-wave mixing [21], were also theoretically predicted with practical considerations for experiment.

For a multi-level electromagnetically induced transparency (EIT) [22, 23] atomic system, other than modifying the linear absorption and dispersion properties for the probe laser beam, the atomic coherence induced by the coupling and probe beams also greatly enhances the strength of near-resonant nonlinear interaction, and the Kerr-nonlinear index of refraction can be modified and significantly enhanced near the two-photon atomic resonance.[24, 25] Such enhanced nonlinearity produces certain unavoidable effects during the self-imaging process and has to be taken into account. The modified Kerr-nonlinear coefficient $n_2$ even changes sign when the sign of coupling or probe frequency detuning is altered, which can actually be used to balance the linear dispersion by considering the definition of the total refractive index $n=n_0+n_2 I_c$ (with $n_0$ and $I_c$ defined as the linear refractive index and the intensity of coupling beam, respectively).[26, 27] This implies that the frequency detunings can certainly exert modulation on the spatially periodic refractive index in the self-imaging process.

Inspired by the promising prospect of the optically induced self-reconstruction or self-imaging effect, we set up an experiment to demonstrate such EITE by forming an EIG under the EIT condition in a three-level $^{85}$Rb atomic configuration. The EIG is achieved by launching a probe field into a spatially distributed optical lattice. The optically induced lattice (along the transverse direction $x$) inside the atomic vapor cell is generated by interfering a pair of coupling laser fields from a same external cavity diode laser (ECDL). As a result, we can observe the spatially intensity-modulated probe field [28] at the output plane of the vapor cell, which implies the formation of the spatially distributed probe-field susceptibility in the atomic medium. The EIT-assisted Talbot effect is manifested through the repetition of the image on the output surface of the cell at the integer Talbot planes. The experimentally measured axial repetition period matches well with the calculated Talbot length of $d^2/\lambda_1$, where $d$ and $\lambda_1$ are defined as the spatial period of the optical lattice and wavelength of the probe field, respectively. Although the current work is implemented in an atomic vapor cell, the experimental demonstration will surely work better in ultracold atoms.

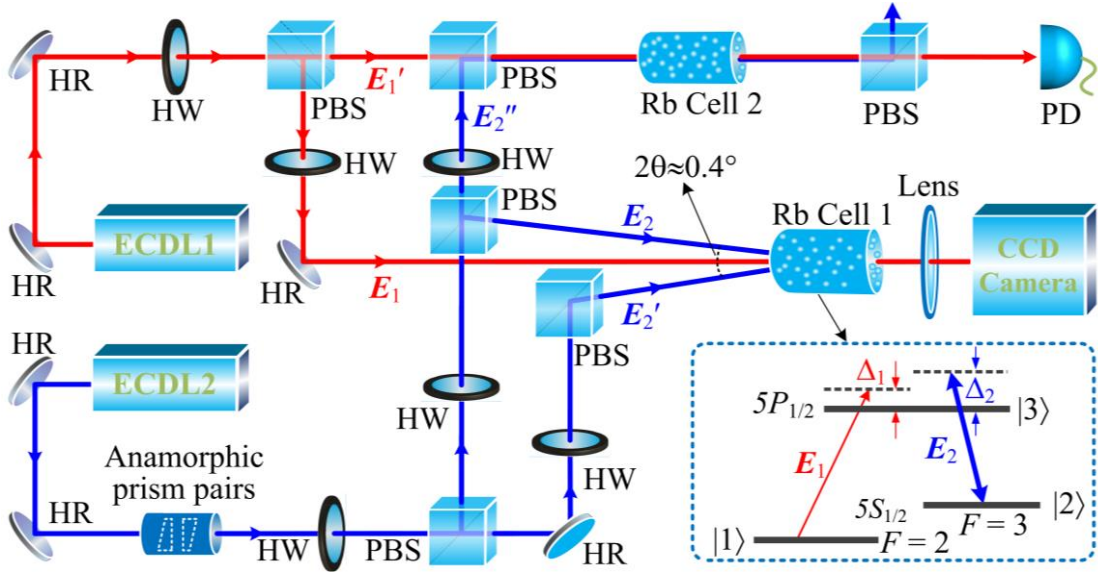

Figure 1 (color online). Experimental setup and the Λ-type energy-level configuration (in the dashed box). The output of the probe beam is imaged onto a CCD camera with a lens. Beams $E_1'$ and $E_2''$ are injected into the auxiliary cell2 to generate an EIT window (detected by PD) in the frequency domain to calibrate the frequencies of the probe and coupling lasers. ECDL: external cavity diode laser, HW: half-wave plate, HR: high reflectivity mirror, PBS: polarization beam splitter, PD: photodiode detector, CCD: charge coupled device.

The experimental setup and relevant energy-level structure are shown in Fig. 1. The probe field $E_1$ (wavelength $\lambda_1$=794.97 nm, frequency $\omega_1$, horizontal polarization, Rabi frequency $\Omega_1$) co-propagates with the two elliptically-shaped coupling beams to interact with the three-level $\Lambda$-type $^{85}$Rb atomic system (the energy-level structure in the dashed box of Fig. 1), which consists of two hyperfine states $F=2$ (state $|1\rangle$) and $F=3$ (state $|2\rangle$) of the ground state $5S_{1/2}$ and one excited state $5P_{1/2}$ (state $|3\rangle$). Two coupling beams $\boldsymbol{E}_2$ and $\boldsymbol{E}_2{'}$ ($\lambda_2$=794.97 nm, $\omega_2$, vertical polarization, $\Omega_2$ and $\Omega_2{'}$, respectively) from a single-mode tunable ECDL2 are symmetrically arranged with respect to the $z$ direction, and intersect at the center of Rb cell1 with an angle of $2\theta\approx0.4°$ to construct an optically induced atomic lattice along the transverse direction $x$ when the frequency detuning $\Delta_2$ is tuned to be near resonance with the transition $|1\rangle\rightarrow|2\rangle$. Here $\Delta_i=\omega_{ij}-\omega_i$ is the detuning between the atomic resonant frequency $\omega_{ij}$ ($j$=1, 2, 3) and the laser frequency $\omega_i$ of $\boldsymbol{E}_i$ ($i$=1, 2). The 7.5 cm long cell1 is wrapped with $\mu$-metal sheets and heated by a heat tape to provide an atomic density of $\sim2.0\times10^{12}$cm$^{-3}$ at 80°C. The spatial periodicity of the optically induced lattice along $x$ direction is calculated to be $d=\lambda_2/2\sin\theta\approx114$ $\mu$m.

With the near-parallel weak probe beam $\boldsymbol{E}_1$ having an elliptical-Gaussian intensity profile from another ECDL1 launched into the induced lattice, the probe-field susceptibility can be spatially modulated under the EIT condition. [28, 29] By carefully choosing the parameters such as frequency detunings and Rabi frequencies of the probe and coupling fields, clear diffracted probe beam pattern from the formed EIG can be observed at the output plane of the cell, which is monitored by utilizing a charge coupled device (CCD) camera (see Fig. 1) with an imaging lens. The Rabi frequency is defined as $\Omega_i=\mu_{ij}E_i/\hbar$ between transition $|i\rangle\leftrightarrow|j\rangle$, where $\mu_{ij}$ is the dipole momentum and $E_i$ is the amplitude of the electric field from $\boldsymbol{E}_i$. The propagation characteristics of the spatially intensity-modulated probe field out of the cell can be imaged onto the CCD camera by moving the lens (which is placed on a precision translation stage) along the $z$ direction. In the meanwhile, the CCD camera (fixed on anther translation stage) is also moved to make the distance (along the $z$ direction) between the camera and the lens be constantly equal to twice of the focal length of the lens, which guarantees the consistency of imaging results at different observing planes. In addition, we use the standard EIT technique to calibrate the frequencies of $\boldsymbol{E}_1$ and $\boldsymbol{E}_2$ fields on the $D1$ transition line, which is realized by coupling two beams $\boldsymbol{E}_1{'}$ and $\boldsymbol{E}_2{''}$ (from the above ECDL1 and ECDL2, respectively) into a second auxiliary cell2 to generate the $\Lambda$-type EIT spectrum for reference.

The key for the current experiment is to periodically modulate the refractive-index profile of the probe field when it propagates through the one-dimensional optical lattice. With the EIT condition satisfied in the Λ-type atomic configuration, the atomic medium modifies the amplitude of probe-field profile, which behaves the same way as an amplitude grating exerting modulation on an electromagnetic wave and can be expressed as an EIG effect. The Kerr-nonlinear coefficient, expressed as $n_2 \propto \text{Re}[\chi^{(3)}]$ [24], is inevitably modified in the EIT window. As a result, the susceptibility (defined as $\chi=(2N\mu_{31}/\varepsilon_0 E_1)\times \rho_{31}$) between states $|1\rangle$ and $|3\rangle$ is given by $\chi=\chi^{(1)}+\chi^{(3)}(|\Omega_2|^2+|\Omega_2'|^2)$, where the linear $\chi^{(1)}$ and nonlinear $\chi^{(3)}$ susceptibilities are expressed, respectively, as

$$\chi^{(1)} = \frac{iN|\mu_{31}|^2}{\hbar\varepsilon_0} \times \frac{1}{\left(\Gamma_{31}+i\Delta_1+\dfrac{|\Omega_2|^2+|\Omega_2'|^2+2\Omega_2\Omega_2'\cos(2k_2 x)}{\Gamma_{21}+i\Delta_2}\right)}, \quad (1)$$

and

$$\chi^{(3)} = \frac{-iN|\mu_{31}|^2}{\hbar\varepsilon_0} \times \frac{1}{\left(\Gamma_{31}+i\Delta_1+\dfrac{|\Omega_2|^2+|\Omega_2'|^2+2\Omega_2\Omega_2'\cos(2k_2 x)}{\Gamma_{21}+i\Delta_2}\right)^2 \times (\Gamma_{21}+i\Delta_2)}. \quad (2)$$

In the above expressions, $\Gamma_{ij}=(\Gamma_i+\Gamma_j)/2$ is the decoherence rate between states $|i\rangle$ and $|j\rangle$; $\Gamma_i$ is the transverse relaxation rate determined by the longitudinal relaxation time and the reversible transverse relaxation time; $N$ is the atomic density at the ground-state $|1\rangle$. Furthermore, by adopting the plane-wave expansion method, the one-dimensional periodically-modulated total refractive index [30, 31] can be described as

$$n(x) = n_0 + \Delta n_1 \cos(2k_2 x) + \Delta n_2 \cos(4k_2 x), \quad (3)$$

where $n_0$ is the uniform refractive index independent of the spatial periodicity; $\Delta n_1$ and $\Delta n_2$ with different spatial periodicities are the coefficients for spatially varying terms in total refractive indices, respectively. Here $n_0$ contains the linear and periodicity-independent parts contributed by $\chi^{(1)}$ and $\chi^{(3)}$, both of which also contribute to $\Delta n_1$; and the term $\Delta n_2 \cos(4k_2 x)$ accounts for the nonlinear index (from $\chi^{(3)}$) inside the grating. Equation (3) shows the physical picture for the spatially modulated refractive index experienced by the probe field with the presence of the standing-wave coupling field. In fact, mathematically, the spatial distribution of the total refractive index (containing terms $\cos(2k_2 x)$ and $\cos(4k_2 x)$) can be shifted/modulated along the transverse direction $x$ under certain parameters, which can be viewed as a result from the balance between the linear and nonlinear refractive indices. Details are provided in the Supplementary Material. To precisely show the EITE with nonlinearity, theoretical simulations are also presented based on the theoretical proposal [17].

The transmission of the modulated probe field at the output surface of the cell is

$$T(x,L) = T(x,0)\exp(-k_1\chi''L/2 + ik_1\chi'L/2), \tag{4}$$

where $\chi'$ and $\chi''$ are the real and imaginary parts of the susceptibility $\chi$, respectively; $T(x,0)$ is the input profile of plane-wave $E_1$; $L$ is the length of the cell.

Considering the periodicity of $\chi = \chi' + i\chi''$, the representation of probe transmission can be rewritten in the form of Fourier optics. According to the Fresnel-Kirchhoff diffraction theory and under the paraxial approximation [3, 17], the amplitude of the probe field at the observation plane with a distance $Z$ from the output surface of the cell can be expressed as

$$T(X,Z) \propto \int_{-\infty}^{+\infty} T(x,L)\exp\left[ik_1\left(Z + \frac{\xi^2}{2Z} - \frac{\xi X}{Z} + \frac{X^2}{2Z}\right)\right]d\xi, \tag{5}$$

where $\zeta$ and $X$ represent the coordinates of the object and observation plane, respectively. Completing the integral in Eq. (5) with the Fourier series expansion of $T(x, L)$, we can obtain the conventional Talbot effect as described by [17]

$$T(x,Z) \propto \sum_{n=-\infty}^{+\infty} C_n \exp\left(-i\pi\lambda_1 n^2 Z/d^2 + i2\pi nX/d\right). \tag{6}$$

According to Eq. (6), the probe transmission with a distance of $Z_T$ ($Z_T = md^2/\lambda_1$ is the Talbot length with the positive integer $m$ viewed as the self-imaging number) can repeat the amplitude at the output plane of the vapor cell with and without shifted half period $d/2$ for odd and even integers $m$, respectively, which is in accordance with the traditional Talbot effect. [3] With $d = \lambda_2/2\sin\theta$ and $\lambda_1 \approx \lambda_2$ considered in our experimental setup, we have

$$Z_T = md^2/\lambda_1 = m \times \left(\frac{\lambda_2}{2\sin\theta}\right)^2 \times \frac{1}{\lambda_1} \approx m \times \frac{\lambda_2}{(2\sin\theta)^2} = m \times 1.63\,cm. \tag{7}$$

By considering $\chi = \chi^{(1)} + \chi^{(3)}(|\Omega_2|^2 + |\Omega_2'|^2)$, the numerical plot for the Talbot effect is shown in Fig. 2(a). Figure 2(b) is the intensity distributions at different Talbot distances during the self-imaging process corresponding to Fig. 2(a). By comparing Fig. 2(b) with the situation when only $\chi^{(1)}$ is considered, we find that the Talbot length is the same for both cases, which agrees with the prediction that the Talbot distance is determined by the periodicity of the induced optical lattice. The effects introduced by $\chi^{(3)}$ are reflected from at least two aspects. One is that third-order nonlinearity can render the change on the intensity profiles of the generated EIG as well as the intensity of the diffracted probe near the resonance. Another one is the introduction of $\chi^{(3)}$ can shift the output probe half period along the $x$ direction when $\Delta_1$ is tuned across a certain value, while such shift is not obtained in the linear case.

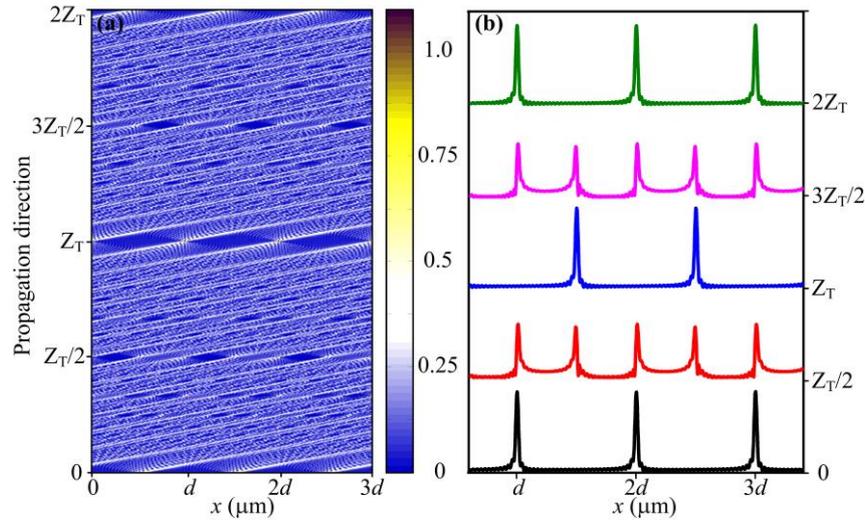

Figure 2 (color online). (a) The calculated Talbot effect carpet for the Λ-type atomic configuration with the third-order nonlinearity considered. The parameters are $\Omega_1=\Gamma_{10}$, $\Omega_2=\Omega_2'=20\Gamma_{10}$, $\Delta_1=10$ MHz, and $\Delta_2=0$. (b) The calculated intensity distributions of the probe field at the output surface (Z=0), $Z=Z_T/2$, $Z=Z_T$ and $Z=2Z_T$, respectively.

Similar to the theoretical derivation, the most important ingredient for the experiment is a well-established periodic susceptibility (namely, the optically induced atomic lattice). With the probe beam (Fig. 3(a)) launched into the lattice constructed by the interference of two coupling beams, we can observe a clear periodic probe beam pattern diffracted by the formed EIG as shown in Fig. 3(c), where the EIT window is generated under the two-photon Doppler-free condition [23], so the Doppler effect plays a less important role here. The generated interference fringe pattern in a 3-dimensional view is depicted in Fig. 3(b). Figure 3(d) is the observed spectra of the probe field from the auxiliary Rb cell2 with the lower and upper curves representing the absorption spectra without and with the EIT window (or coupling beam), respectively. Note that optical pumping effect [24] exists simultaneously with the EIT in such Λ-type system, which increases the absorption of the probe field on *D*1 line and makes the intensity of the probe transmission at $\Delta_1+\Delta_2=0$ a little weaker than the case with $\boldsymbol{E}_2$ turned off.

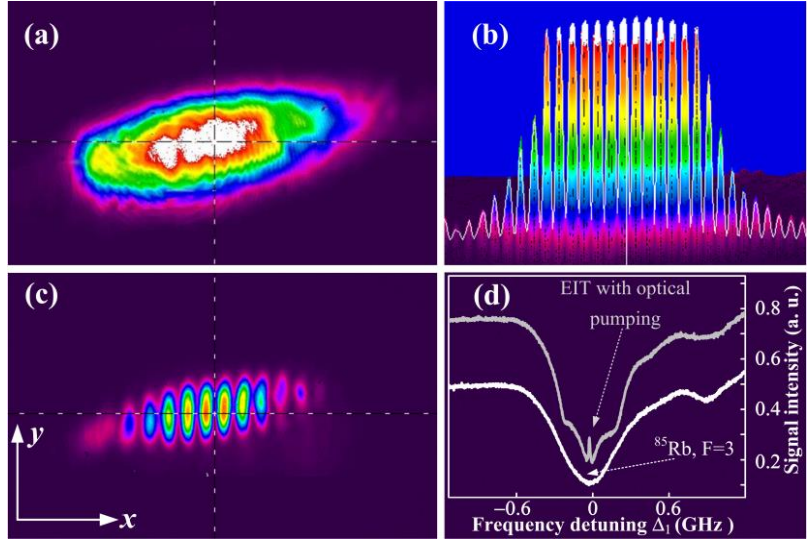

Figure 3 (color online). (a) Image of the probe beam. (b) Observed 3D-view interference pattern of the two coupling beams. (c) The experimentally generated diffraction pattern on the probe field. (d) The reference EIT signal from an auxiliary cell. The upper and lower curves are the generated EIT signal and the absorption spectrum corresponding to the transition of $^{85}$Rb, $F=3 \rightarrow F'$, respectively.

With the EIG-diffracted image observed at the output plane of the cell with a coordinate of $\zeta=L$, we monitor the propagation characteristic of the intensity-modulated probe beam at different observation planes. According to the previous description, the CCD camera and the imaging lens move simultaneously with the same distance along the $z$ direction to maintain the space between them unchanged. The observed images at different $z$ planes are shown in Figs. 4(a1)-(a9). According to Eq. (7), the Talbot distance is calculated to be $m\times1.63$ cm. The experimental results show that the diffracted image shifts half period (compared to that at Z=1.5) when the observation plane is tuned to be at Z≈1.5 cm, which agrees with the theoretically calculated Talbot distance of $Z=Z_T=1.63$ cm. When the observation coordinate exceeds Z=1.5 cm (the first Talbot plane, $m$=1), the diffraction image shifts back (along the opposite direction to the movement on the first Talbot plane) to match the image at $\zeta=L$. Due to the limitation on the maximum displacement (2.5 cm) of the precision translation stage, the image at the second Talbot plane ($m$=2) is not observed. However, the moving trend of the images at Z=1.8 cm, 2.1 cm and 2.5 cm indicates that the image at Z≈3 cm should match the one at the output surface of the cell ($\zeta=L$) without a spatial shift The experimental images at the first ($m$ is an odd integer) and second ($m$ is an even integer) Talbot planes support the theoretical predictions [17], as indicated in Fig. 2(a). Also, Fig. 4(b) shows the experimentally measured evolution of Talbot length with a controlled periodicity of the interference fringes by slightly varying the angle $2\theta$ between the coupling beams $E_2$ and $E_2'$. For a given angle $\theta$, there exist two experimental squares for each angle, which represent the maximum and minimum values of Z for which the measured diffraction patterns show self-reconstruction. The experimental results basically agree with the theoretical (solid) curve, which indicates that the Talbot length can be controlled by the period of the induced optical lattice. Furthermore, fractional Talbot effect has also been observed as shown in Fig. 5, where the periodicity of the diffraction pattern can be doubled along the *x* direction at $Z=Z_T/2$ and $Z=3Z_T/2$. Based on the measured integer Talbot effect, the Talbot length $Z_T$ is approximately 15 mm. The periodicity-doubling images were observed around Z=8 mm ($\approx Z_T/2$) and Z=22 mm ($\approx 3Z_T/2$), respectively, which match well the calculated fractional Talbot length, as shown in Fig. 2(b).

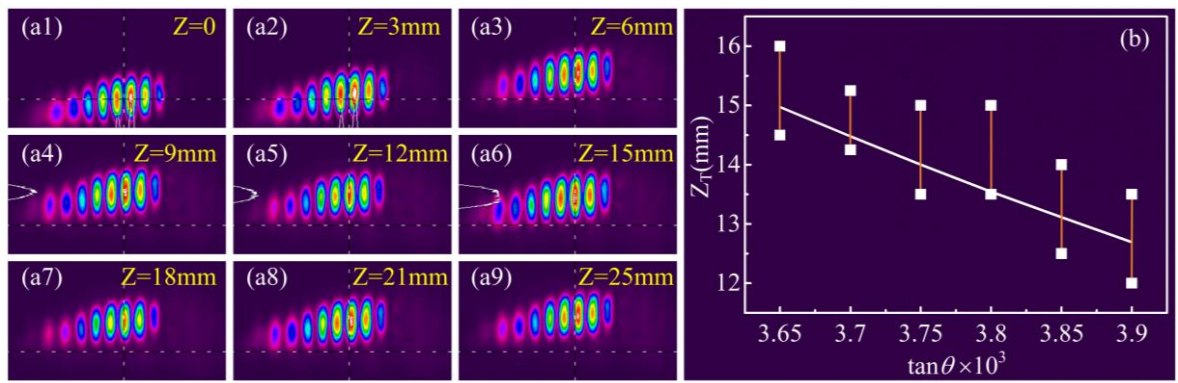

Figure 4 (color online). Demonstration of the electromagnetically induced Talbot effect. With the observation plane moving along the *z* direction, the intensity profile of the probe field at $Z=mZ_T$ matches the profile at the output plane of cell with (*m* is an odd integer) and without (*m* is an even integer) shifting half a period. (j) Dependence of the Talbot length on the angle between the two coupling beams. The squares are the experimental observations and the solid curve is the theoretical prediction.

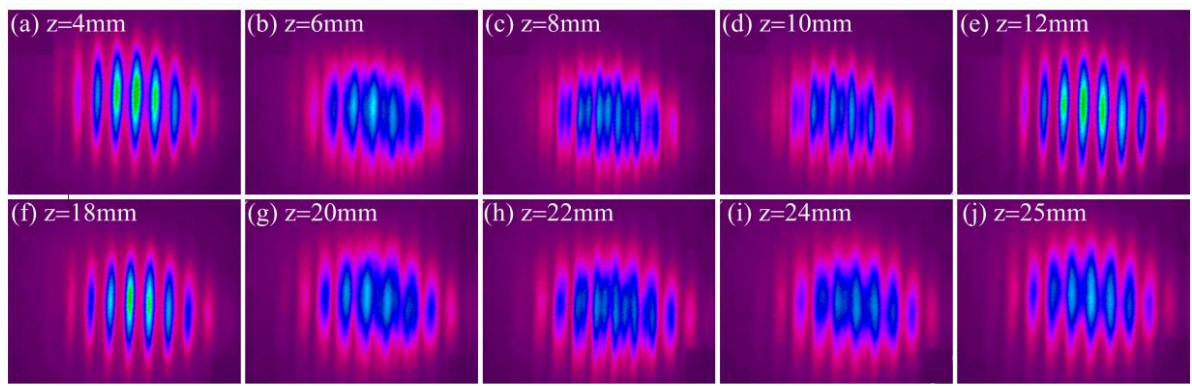

Figure 5 (color online). Observed fractional Talbot effect (double periodicity) at Z=8 mm ($\approx Z_T/2$) and Z=22 mm ($\approx 3Z_T/2$), respectively. The experimentally observed integer Talbot length is at $Z_T \approx 15$ mm.

In experiment, the Kerr-type nonlinearity in the atomic medium with periodically modulated indices is verified by observing a spatial shift (induced by the combined effects of the linear and nonlinear dispersions near resonance) of the diffracted probe. As shown in Fig.

6, with the frequency detuning of the coupling field fixed at $\Delta_2=0$MHz, the diffracted intensity pattern from the generated EIG can vary with the two-photon detuning $\delta=\Delta_1-\Delta_2$. The periodical probe beam has the strongest value near the resonance $\delta=\Delta_1-\Delta_2=0$, while it becomes the weakest at the points far away from resonance ($\delta=-30$ MHz and 40 MHz). Particularly, the diffracted pattern shifts a $d/2$ distance along the transverse $x$ direction when the frequency detuning varies between $\Delta_1=0$ to $\Delta_1=10$MHz. Such observed shift of the spatially intensity-modulated probe beam in the EIT medium can be attributed to the abrupt sign jump (negative↔positive) of the cross-Kerr nonlinear index $n_2$ in the frequency detuning range [32]. To be more specific, both terms $\Delta n_1$ and $\Delta n_2$ in Eq. (3) can be manipulated by controlling the frequency detuning $\Delta_1$ of the probe field according to the expressions in Eqs. (1) and (2). As a result, when $\Delta n_1$ and $\Delta n_2$ are in the same magnitude level [33] by properly settling the parameters, the total $n(x)$ from the sum of $n_0$, $\Delta n_1\cos(2k_2x)$ and $\Delta n_2\cos(4k_2x)$ can have a spatial shift of a half period when $n_2$ changes from negative to positive, which can also force the sign saltation of $n_2$. Such spatial shift of the diffracted probe pattern can be used to provide another way of modulating the imaging process in ultracold atomic clouds without modifying the experimental setup.

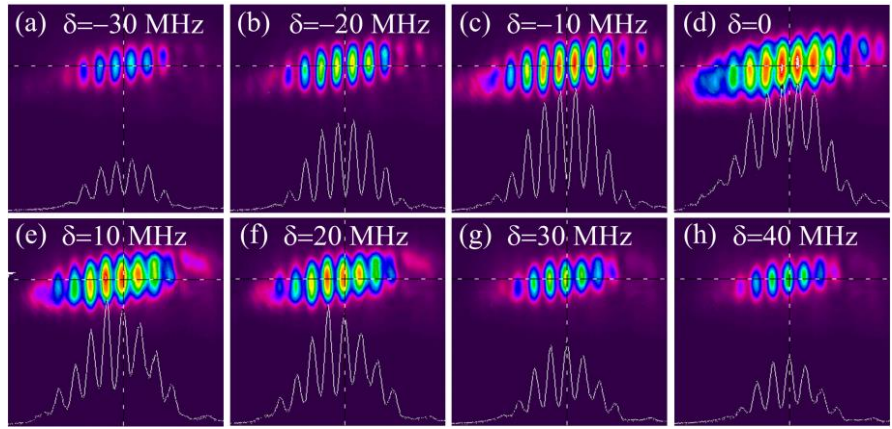

Figure 6 (color online). Experimentally observed diffraction probe beam patterns by the formed EIG versus the two-photon detuning $\delta=\Delta_1-\Delta_2$. (a) $\delta=-30$ MHz, (b) $\delta=-20$ MHz, (c) $\delta=-10$ MHz, (d) $\delta=0$, (e) $\delta=10$ MHz, (f) $\delta=20$ MHz, (g) $\delta=30$ MHz and (h) $\delta=40$ MHz, respectively.

In summary, the demonstrated EIT-assisted Talbot effect in a three-level atomic system can provide a new and powerful methodology to image ultracold atomic or molecular clouds. Such a lensless self-imaging technique may be less affected by various vibrations during practical applications. In order to simplify the experimental observation, a secondary imaging process is implemented by utilizing a convex lens with the right focal length, which doesn't conflict with the proposed lensless imaging scheme. Owing to the easy controllability of the linear absorption/dispersion properties and the Kerr nonlinearity, such coherently-prepared multi-level atomic systems can be used as an ideal platform to further investigate intriguing nonlinear/quantum beam dynamical features predicted for constructed periodic optical systems beyond the simple Talbot effect. For example, our system can be promisingly applied to demonstrate the proposed parity-time-symmetric Talbot effect [7] by adding another standing-wave pump field to construct parity-time-symmetric potential [29]. Although the current EITE experiment was performed in an atomic vapor cell, the experimental demonstration can certainly be extended to ultracold atomic clouds, which can have profound implications in exploring interesting properties of BEC and atomic lattices in ultracold atomic and molecular clouds, as well as in quantum information science.